# Error Corrected Spin-State Readout in a Nanodiamond

Jeffrey Holzgrafe[1,2,]*, Jan Beitner[1,]*, Dhiren Kara[1], Helena S. Knowles[1,3], and Mete Atatüre[1,†]

[1] *Cavendish Laboratory, University of Cambridge, JJ Thomson Avenue, Cambridge CB3 0HE, UK*

[2] *John A. Paulson School of Engineering and Applied Sciences, Harvard University 29 Oxford Street, Cambridge MA 02138, USA*

[3] *Department of Physics, Harvard University, 17 Oxford Street, Cambridge MA 02138, USA*

* These authors contributed equally to this work

† Corresponding author M.A. (ma424@cam.ac.uk)

Quantum state readout is a key component of quantum technologies, including applications in sensing, computation, and secure communication. Readout fidelity can be enhanced by repeating readouts. However, the number of repeated readouts is limited by measurement backaction, which changes the quantum state that is measured. This detrimental effect can be overcome by storing the quantum state in an ancilla qubit, chosen to be robust against measurement backaction and to allow error correction. Here, we protect the electronic-spin state of a diamond nitrogen-vacancy center from measurement backaction using a robust multilevel $^{14}$N nuclear spin memory and perform repetitive readout, as demonstrated in previous work on bulk diamond devices. We achieve additional protection using error correction based on the quantum logic of coherent feedback to reverse measurement backaction. The repetitive spin readout scheme provides a 13-fold enhancement of readout fidelity over conventional readout and the error correction a 2-fold improvement in the signal. These experiments demonstrate full quantum control of a nitrogen-vacancy center electronic spin coupled to its host $^{14}$N nuclear spin inside a ~25 nm nanodiamond, creating a sensitive and biologically compatible platform for nanoscale quantum sensing. Our error-corrected repetitive readout scheme is particularly useful for quadrupolar nuclear magnetic resonance imaging in the low magnetic field regime where conventional repetitive readout suffers from strong measurement backaction. More broadly, methods for correcting longitudinal (bit-flip) errors described here could be used to improve quantum algorithms that require nonvolatile local memory, such as correlation spectroscopy measurements for high resolution sensing.

# Introduction

Quantum devices rely on the readout of a quantum system after a period of controlled evolution or interaction with other systems. High readout fidelity is critical to such devices, because it enables heralded state initialization[1–3], quantum error correction[4], and improved sensitivity of quantum sensors[5,6]. Weak signals are typical for quantum systems and limit the information that can be gained about the state of the system in one readout step. In addition, all quantum measurements exhibit backaction, which changes the device's state during readout. In the absence of backaction, namely for an ideal quantum non-demolition measurement, the problem of a weak signal could be overcome by multiple repeated readout steps. In this article, we describe and demonstrate an error correction protocol that protects spin memories from measurement backaction in the form of depolarization due to spin-spin interactions.

We choose to implement our protocol in a nitrogen-vacancy center (NV) in diamond, a versatile system for quantum sensing, due to the ease of controllability of its electronic spin and the in-built multi-level local nuclear memory: the three spin states of the nitrogen-14 nuclear spin (see illustration in Fig. 1a). The NV center has emerged as a promising candidate for quantum information[7,8] and sensing[6,9–11] applications because its electronic spin has a long coherence time at room temperature and can be easily initialized, readout and controlled with optical and microwave (MW) fields. However, room-temperature optical readout using the NV center's spin-dependent fluorescence yields only $\approx 0.02$ detected photons in typical experiments like ours before the spin state is repolarized, leading to a low readout fidelity. Recent experiments on bulk-diamond NV centers overcome this poor readout by using nearby hyperfine-coupled nuclear spins as memory qubits to store and repetitively readout the NV center's electronic-spin state[3,5,12–16]. Other techniques[17] – predominantly excitation of optical cycling transitions[18,19] – can also improve readout fidelity, but such techniques are not widely used in room-temperature sensing experiments as they require additional lasers, cryogenic temperatures, or sensitive electrical measurements. The fidelity achievable with the nuclear-spin repetitive readout technique is limited by measurement backaction in the form of nuclear spin depolarization that occurs when the NV center is optically excited. Here we show that our error correction protocol can partially reverse this depolarization and improve readout fidelity by using the NV electronic spin as an ancilla qubit to control the host [14]N nuclear spin. Crucially, our protocol does not require measurement, which has low fidelity for our system, because we use the quantum logic of coherent feedback[20,21] to perform error correction. We demonstrate that this protocol can enhance the repetitive readout fidelity of a nitrogen-vacancy (NV) center electronic spin by more than 50% in the regime where backaction is strong.

We apply our protocol to an NV center in a high-purity nanodiamond (Fig. 1a) with typical diameter of 25 nm[22]. Building upon our recent demonstration of preparation, readout and coherent control of a [13]C nuclear spin in a nanodiamond[23], here we demonstrate repetitive readout of a [14]N nuclear spin in a nanodiamond. These techniques, which have been widely used in bulk NV center experiments, enable a wide range of nuclear-spin assisted measurements and enhanced sensitivity[5,13,24]. By combining nuclear spin-assisted readout with the small size of these nanodiamonds we hope to enhance NV sensors for applications such as scanning magnetometry, interferometry of mechanical motion[25], and intracellular nanoscale measurements[26,27].

# Results

**Coherent Control and Repetitive Readout of a Host [14]N nuclear spin in a Nanodiamond**

Figure 1b presents the NV electronic spin ($S = 1$) and the [14]N nuclear spin ($I = 1$) hyperfine levels in the ground-state manifold, including the electronic microwave (MW) and nuclear radio frequency (RF) transitions, and the optically induced spin flip-flop transitions, $\gamma_\pm$ that occur between the electronic and nuclear spin in the electronic excited state. In order to drive direct coherent rotations that target individual hyperfine energy levels, NV center electronic-spin coherence times $T_2^* \gtrsim 1$ μs are required so that the hyperfine transitions are well resolved[7,28,29]. Here, we are able to reach this regime by using high-purity nanodiamond crystals with a relatively low nitrogen defect concentration of 50 ppm[22].

Figure 1c shows three well-resolved dips in the optically detected magnetic resonance (ODMR) signal (black curve) resulting from the nuclear-spin selective electronic spin transitions (MWB, MWC, and MWE). The dip amplitudes of these three transitions correspond to the nuclear spin population, providing a direct measure of nuclear-spin polarization. The middle green ODMR curve of Fig. 1c, taken with a magnetic field $B_0 = 51$ mT aligned with the NV crystal axis, shows the nuclear spin is polarized into the state $|+1\rangle_n$ with high fidelity ($> 95\%$). This nuclear spin polarization is caused by the combination of fast spin flip-flop transitions $\gamma_-$ and the optical pumping of the electronic-spin into $|0\rangle_e$, illustrated by the solid green wavy and straight arrows, respectively, in Fig. 1b. The $\gamma_\pm$ transitions are allowed due to the strong perpendicular components of the hyperfine interaction in the NV center's excited state, which induce spin-mixing[30]. In a single optical cycle, the spin flip-flop probabilities $\gamma_\pm/\gamma_o$ for each transition are

$$\frac{\gamma_\pm}{\gamma_o} = \frac{2A_{es}^2}{2A_{es}^2 + (D_{es} \pm g_e \mu_B B_0)^2}, \quad (1)$$

where $\gamma_o$ is the optical pumping rate induced by laser excitation, $g_e \approx 2$ is the electron g-factor, $\mu_B$ is the Bohr magneton, and $A_{es}$ and $D_{es}$ are the perpendicular component of the hyperfine interaction and the electronic-spin zero-field energy splitting of the NV center's optical excited state, respectively[31,32]. The $\gamma_\pm$ transitions are strongest near the excited state level avoided crossing (ESLAC) at $B_0 = \mp B_{ESLAC} = \mp D_{es}/(g_e \mu_B) = \mp 51$ mT due to the strong spin mixing (Fig. 1d inset). In the moderate-field regime where $B_0 \approx B_{ESLAC}$, the $\gamma_-$ transitions are $10^3$ times stronger than the $\gamma_+$ transitions, as shown in Fig. 1d. This imbalance between $\gamma_+$ and $\gamma_-$ together with the optical electronic spin polarization into $|0\rangle_e$ creates the dynamic nuclear polarization seen in the middle green curve of Fig. 1c. In contrast, in the low-field regime ($B_0 \ll B_{\text{ESLAC}}$), $\gamma_+$ and $\gamma_-$ are comparable, and polarization does not occur, as shown in the top black curve of Fig. 1c.

In the high-field regime where $B_0 \gg B_{\text{ESLAC}}$ the $\gamma_\pm$ transition strengths have similar magnitude and are much weaker overall, which reduces the steady state degree of polarization and the rate of polarization. To polarize the nuclear spin in this regime we transfer electron spin polarization onto the nuclear spin using repeated SWAP operations, which exchange the electronic and nuclear spin states and are created by direct RF rotations (see Supplementary Note 1) of the nuclear spin and hyperfine selective MW pulses[31,33]. Figure 1c shows 85% polarization into $|0\rangle_n$ reached at 244 mT (bottom blue curve), indicating that a high degree of nuclear spin polarization can be reached in nanodiamond-hosted NV centers in the high-field regime.

In this high-field regime, the slow $\gamma_\pm$ flip-flop transitions make it possible to perform repetitive readout of the nuclear spin state with weak measurement backaction[3], as described earlier. We use the protocol shown in Fig. 2a to characterize repetitive readout of the NV center's electronic spin state after it has been stored in the nuclear spin. After initializing the system into the state $|0\rangle_e|+1\rangle_n$, we use the MWB transition (see Fig. 1b) to either prepare the electronic spin state in $|-1\rangle_e$ or leave it in $|0\rangle_e$. A CNOT gate created by two RF pulses maps the electronic spin state onto a nuclear spin state so that $|0\rangle_e|+1\rangle_n \to |0\rangle_e|+1\rangle_n$ and $|-1\rangle_e|+1\rangle_n \to |-1\rangle_e|-1\rangle_n$. Finally, the nuclear spin state is measured by repetitive readout of the ODMR on the MWE transition. We store the original electronic spin state in the $\{|\pm 1\rangle_n\}$ code space because this mapping provides a slow loss of relative polarization due to the buffer $|0\rangle_n$ state[3]. We quantify the repetitive readout by the total number of detector counts, $C_0$ and $C_1$, that are collected when the electronic spin is initialized into the $|0\rangle_e$ or $|-1\rangle_e$ state, respectively. The figure of merit for this readout scheme is the readout fidelity[6,34],

$$F = \left(1 + \frac{(C_0+C_1)}{(C_0-C_1)^2}\right)^{-\frac{1}{2}}. \tag{2}$$

This takes into account both shot noise and spin-projection noise and quantifies the efficiency of a readout scheme, approaching unity for an ideal projection-limited readout.

As shown in Fig. 2b, the repetitive readout protocol can improve readout fidelity by more than a factor of 10 from $F = 0.03$ for a single electronic-spin readout to $F = 0.4$ for $N = 2300$ repetitive nuclear spin readouts. The cumulative signal $C_0 - C_1$ increases linearly with repeated readout steps initially but saturates exponentially with a characteristic scale $N_{1/e} \sim 1700$ (Fig. 2b inset). This saturation occurs because information about the nuclear spin state is lost to measurement backaction induced by the $\gamma_\pm$ transitions. As the cumulative signal saturates and the shot noise increases with $\sqrt{N}$, the readout fidelity decreases for large numbers of readouts as shown in the main plot of Fig. 2b.

These results are consistent with a model that includes only the depolarization errors due to the $\gamma_\pm$ flip-flop transitions during optical readout without any additional depolarization mechanisms (Supplementary Note 2). We thus predict that repetitive readout in our nanodiamonds is limited only by this flip-flop interaction occurring in the electronic excited state, and that working in an even higher-field regime ($B_0 \gg B_{\text{ESLAC}}$, with reduced $\gamma_\pm$ rates) would lead to additional enhancements in repetitive readout fidelity, including single-shot readout[3].

**Error-Corrected Repetitive Readout**

We now present an error correction protocol which can improve repetitive readout fidelity by correcting errors caused by the $\gamma_\pm$ transitions. We take advantage of the error imbalance between the weak $\gamma_+$ and strong $\gamma_-$ transitions in the moderate-field regime (Fig. 1d) to design a classical error correction protocol using the three nuclear-spin states as our logical storage space. The frequent $\gamma_-$ type flip-flop errors from $|-1\rangle_n$ to $|0\rangle_n$ can be corrected by periodically performing coherent feedback to transfer any population residing in $|0\rangle_n$ back into the $|-1\rangle_n$ state (see illustration in Fig. 3a, top panel). The feedback procedure may actually create an additional error if a $\gamma_+$ type flip-flop from $|+1\rangle_n$ to $|0\rangle_n$ has occurred, but this type of flip-flop error is rare because $\gamma_+$ is weak.

We implement this protocol using the pulse sequence shown in Fig. 3a, bottom panel. This protocol follows that of conventional repetitive readout, but after an error-correction period of $N_r$ nuclear-spin readouts, we use a combination of a MW and a RF pulse to create a SWAP gate in the relevant subspace of $\{|0\rangle_e, |-1\rangle_e\} \otimes \{|0\rangle_n, |-1\rangle_n\}$. This coherent feedback operation can be understood by considering a density matrix $\rho_0 \propto |0\rangle_e\langle 0|_e \otimes (|-1\rangle_n\langle -1|_n + \epsilon|0\rangle_n\langle 0|_n) + \mathcal{O}(\epsilon^2)$, an approximate representation of the system state after an error-correction period of $N_r$ readout steps, where $\epsilon$ is the probability of memory qubit incrementing error rate after these readouts. If small, this error probability can be approximated as $\epsilon \approx \gamma_- t_r N_r$, where $t_r$ is the laser excitation time used for electron spin readout. As long as the error-correction period is short enough to keep this error probability low, such an error can be corrected. An ideal SWAP gate transforms the density matrix into $\rho \propto |-1\rangle_n\langle -1|_n \otimes (|0\rangle_e\langle 0|_e + \epsilon|-1\rangle_e\langle -1|_e) + \mathcal{O}(\epsilon^2)$. The effect of the SWAP gate can be understood as a transfer of the entropy of the nuclear spin state onto the electronic spin. After this transfer, the entropy on the electronic spin could be pumped away by repolarizing the electronic spin with a dedicated laser pulse after every SWAP gate. However, such a laser pulse would drive the $\gamma_\pm$ transitions and thus induce nuclear-spin depolarization and reduce performance. To avoid this additional depolarization, our protocol instead relies on subsequent readouts to repolarize the electronic spin after the SWAP gate.

This combination of $N_r$ readout steps and one error correction sequence is repeated until $N$ total readouts have been performed. Fig. 3c shows a comparison of the readout fidelity with and without error correction when working in the moderate-field regime ($B_0 = 82$ mT) and using frequent error correction ($N_r = 5$). Without error correction, the repetitive readout fidelity reaches a maximum of $F = 0.08$ for $N = 120$ readouts, much lower than the high-field fidelity presented earlier due to the strong flip-flop error rate $\gamma_-$ in this moderate-field regime. The error correction protocol improves repetitive readout fidelity by $57 \pm 8\%$ to $F = 0.13$ for $N = 235$ readouts. This improvement is equivalent to increasing the NV center brightness by $250 \pm 25\%$. As shown in Fig. 3c, decreasing the error-correction period $N_r$ improves readout fidelity because lower repumping periods reduce the probability of an uncorrectable double-error ($|-1\rangle_n \to |0\rangle_n \to |+1\rangle_n$) occurring between error correction sequences. The improvement in readout fidelity scales as $N_r^{-1/2}$ for small $N_r$. This scaling is an intrinsic feature of the protocol and can be explained by a simple model that assumes ideal error correction operations (Supplementary Note 2).

The solid curve in Fig. 3c shows the results of a master-equation model of the population dynamics. This model includes several non-idealities of our system, such as the finite probability for pulse errors and driving of off-resonant spin transitions (Supplementary Note 3). The only free parameter in the fit to the data in Fig. 3c is the electronic-spin polarization fidelity, which we extract as $P_{|0\rangle_e} = 0.81$, consistent with previous studies[12,35,36]. The model shows that this low electronic-spin polarization fidelity reduces the readout fidelity achievable through the error correction protocol by roughly 30% compared to the ideal scenario in which $P_{|0\rangle_e} = 1$. This reduced performance is due to the finite probability of the system occupying the $|-1\rangle_e|-1\rangle_n$ state before the error correction operation, which causes the SWAP gate to induce an error by transferring the system into the $|0\rangle_n$ state. The sub-unity charge state fidelity (ratio of the neutrally charged $NV^0$ and the negatively charged $NV^-$ population after optical excitation), estimated at 0.75 from our nuclear spin Rabi oscillation data[12], also reduces the corrected readout fidelity by roughly 10% because the error correction sequence has no effect in the $NV^0$ state. The model further predicts an improvement in readout fidelity of around 100% over conventional repetitive readout for $N_r = 1$.

We also perform error-corrected repetitive readout measurements in the high-field regime at $B_0 = 244$ mT and observed a 7% improvement in readout fidelity due to error correction (Supplementary Note 4). The smaller improvement is expected as our scheme targets only $\gamma_-$ type errors and thus operates most effectively in the moderate-field regime where the error imbalance $\frac{\gamma_-}{\gamma_+}$ is large. Similar error correction protocols with a larger logical storage space would enable the correction of more types errors.

Using the model described above, we can quantitatively predict the performance of the protocol under different magnetic field conditions. To do so, we fit Equation 1 to the measured depolarization rates $\gamma_+$ and $\gamma_-$ (Supplementary Note 3) to estimate those rates as a function of the applied magnetic field $B_0$. For simplicity, we assume all other parameters in the model do not change with applied magnetic field. Figure 4 shows the calculated and measured improvement in readout fidelity from error correction, for different error-correction periods $N_r$. In the low and high magnetic field regimes, the similarity between the depolarization rates $\gamma_+$ and $\gamma_-$ reduces the improvement from error correction as described above. The reduced improvement around $B_0 \approx B_{\text{ESLAC}}$ is caused by the fast error rates caused by strong spin mixing. Excluding this regime near $B_{\text{ESLAC}}$, at moderate fields $B_0 \approx 20$ to $140$ mT the estimated improvement can reach more than 1.5-fold for frequent error-correction.

## Discussion

The improvements in readout fidelity achieved with repetitive readout (in the high-field regime) and error corrected repetitive readout (in the moderate-field regime) come at the cost of a longer readout time. Reaching the optimal readout fidelity in the high-field repetitive readout measurements (Fig. 2b) requires a 3.2 ms long measurement. The error-corrected protocol requires 30 µs per correction sequence, dominated by the RF pulse ring-down time, 20 µs, which could be reduced with improved engineering of the RF-signal delivery system. For long measurements such as $T_1$- or $T_2$-sensing schemes the longer readout length may not be prohibitive. Indeed, nanodiamond-hosted NV centers have most widely been applied to relaxometry measurements that detect changes in the electronic-spin $T_1$, which can exceed 1 ms in nanodiamonds[37–39]. This suggests that practical nanodiamond-NV sensors could benefit from the readout enhancing techniques described here.

We emphasize that our error correction protocol reverses depolarization that is intrinsic to the physics of the NV center and is thus also applicable to NV centers in bulk diamond. The long $T_2$ time ($> 10$ ms) accessible in bulk diamond NV centers means that error correction could improve the sensitivity of NV-based magnetometry measurements. In particular, the techniques described here enable high fidelity repetitive spin readout in the moderate magnetic-field regime, increasing the sensing range of the NV center. Lower magnetic field nuclear magnetic resonance (NMR) detection with NV centers could be used to study nanoscale chemical structures in the limit where quadrupolar[5,40,41] coupling dominates over the Zeeman splitting. Quadrupolar coupling has a typical magnitude of order 1 MHz, corresponding to a strong coupling regime in magnetic fields below about 100 mT. In the strong coupling regime, the energy levels contain rich spatial information about the molecular geometry and motion that is inaccessible in traditional ensemble measurements in the Zeeman regime[5,42]. In addition, lower magnetic fields require less expensive and bulky apparatus for portable measurement technology.

Correction of measurement backaction errors can be extended in several ways: Our protocol uses a pulsed error correction sequence derived from quantum logic. A continuous protocol in which error correction transitions are driven at the same time as readout pulses could provide many of the same improvements without the extension in readout time. Additionally, our protocol used only three states as logical resources, which allows us to correct only one type of error. Other nuclear spin memories with larger dimension, such as a register of hyperfine-coupled [13]C nuclei or the host nucleus of the germanium vacancy center in diamond ($I_{73\text{Ge}} = 9/2$) could see dramatically improved performance with similar protocols.

Here, we overcome measurement backaction by using a simple error-correction code to protect longitudinal information stored in the [14]N memory from bit-flip errors. Such classical error-correction codes may prove useful to other quantum information protocols that rely on quantum-addressable classical memories. As a near-term example, recent quantum correlation spectroscopy[24,43] experiments on NV centers achieve high resolution measurement of nuclear magnetic resonance signals by storing classical information in the $^{14}$N nuclear spin state for long correlation waiting periods. However, to prevent the NV electronic spin from dephasing target spins during the waiting period, the NV must be optically pumped, which reduces the $^{14}$N classical storage time. The error-correction protocol described here could be used to extend the classical storage time, and hence improve spectral resolution of these measurements.

## Methods

We used a homebuilt confocal microscope (NA = 0.9 Nikon air objective) with galvanometer mirror scanning to isolate individual nanodiamond-hosted NV centers. A 532 nm laser was modulated by an AOM (Acousto-Optical Modulator, AA optoelectronics MT80-A1-VIS) to create pulses. The laser intensity was calibrated to the saturation power of the NV center, and 850 ns and 350 ns pulses were used to initialize and readout the electronic spin state, respectively. NV center fluorescence during readout was filtered to select the 600 to 800 nm band, and coupled into a single-mode optical fiber and detected by an APD (Avalanche PhotoDiode, Excelitas SPCM-AQRH-14-FC). The APD output was gated by a microwave switch (MiniCircuits ZASWA-2-50DR+).

Microwave and RF signals were generated by a Tektronix 70002A 10 GHz AWG (Arbitrary Waveform Generator) on separate channels, then amplified (using Mini-Circuits ZHL-16W-43+ or ZX60-14012L+ microwave amplifiers and LZY-22+ RF amplifiers), combined (Microwave Circuits D02G18G1 diplexer) and delivered to the sample through a 100 μm diameter single-loop inductor. This inductor was fabricated by electron beam deposition of 2 nm of titanium and 200 nm of gold through a steel shadow mask, followed by electrodeposition of silver and a protective layer of gold to a total thickness of roughly 3 μm on a 500 μm thick intrinsic silicon substrate mounted on an aluminum heatsink. This system enables us to apply π rotations of the NV electronic spin in less than 4 ns with minimal heating effects. We note however, that nuclear-spin selective microwave transitions are driven with 399 ns π-pulses to minimize the probability to drive nearest neighbor transitions.

We note that in our implementation of the repetitive readout protocol, the NV electronic spin is sometimes left in the $|-1\rangle_e$ state after the CNOT gate, which inverts the usual relationship between the fluorescence signal and the initial electronic-spin state that is being readout. However, this has a small effect on our overall measurement because the electronic spin repolarizes after only a few readout cycles.

The nanodiamonds were purchased from Nabond Technologies Co. and dispersed in a solution of 10 mg of nanodiamond powder to 25 mL of ethanol via sonication. This solution was deposited on the sample using an Omron U22 nebulizer. This method produces a well dispersed distribution of nanodiamond locations and allows individual addressing by the confocal microscope. Previous work has shown these nanodiamonds have a typical diameter of $(23 \pm 7)$ nm[22]. Approximately 1% of the NV centers detected with our confocal microscope show the coherence required to resolve the hyperfine coupling in ODMR measurements ($T_2^* \gtrsim 1$ μs).

The static magnetic field was created by a large axially magnetized (Ø 5 X 4) cm neodymium permanent magnet mounted on a two-axis rotation mount, centered on the sample. This setup can produce fields up to 0.25 T limited by the size of the sample heatsink, which prevents the magnet from being placed closer to the sample. Magnetic field alignment was performed by using ODMR measurements to monitor the diagnostic parameter $f_{\text{MWA}} \pm f_{\text{MWB}}$, which displays an extremum as a function of magnetic field orientation when the field is aligned. The $+$ ($-$) condition is chosen for magnetic fields below (above) the ground-state anti-crossing. Based on our calibration, we estimate that using this method we can reduce the off-axis component of the magnetic field to less than 50 μT.

# Acknowledgements

We thank Gavin Morely for stimulating discussions. We gratefully acknowledge financial support by the Leverhulme Trust Research Project Grant 2013-337, the European Research Council ERC Consolidator Grant Agreement No. 617985 and the Winton Programme for the Physics of Sustainability. H.S.K. acknowledges financial support by St John's College through a Research Fellowship. J.H. acknowledges the UK Marshall Aid Commemoration Commission for financial support.

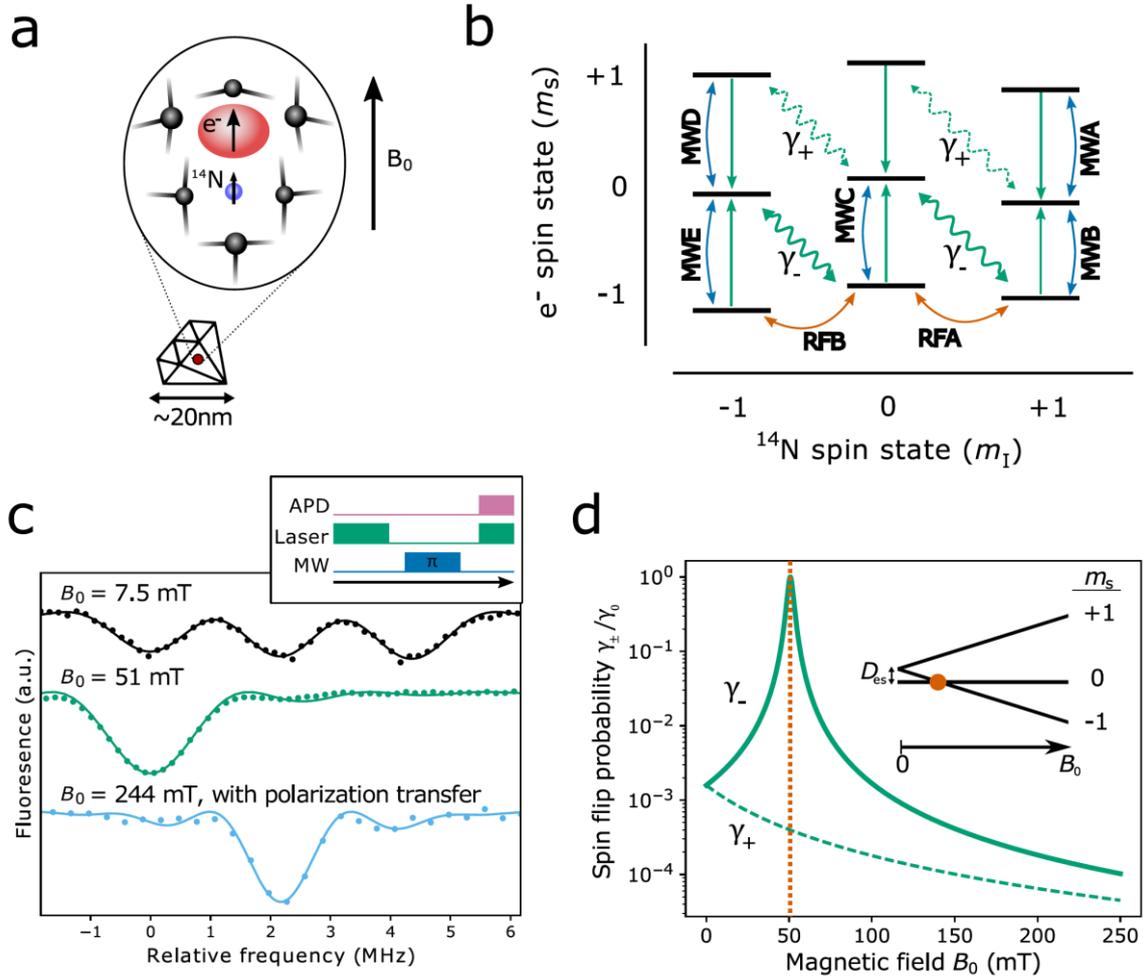

**Figure 1** Initialization and coherent control of a host $^{14}$N nuclear spin in a nanodiamond. **a** Illustration of an NV center in a nanodiamond, with applied magnetic field $B_0$ along the NV axis. **b** Level diagram of the electronic-nuclear spin system in the ground-state manifold in a magnetic field $B_0 = 250\ mT$. Relevant transitions are shown: microwave (MW) electronic-spin transitions (blue arrows), radio-frequency (RF) nuclear-spin transitions (orange arrows), optical electronic-spin polarization (straight green arrows), and the optically-induced spin nuclear-electronic flip-flop transitions (green wavy arrows) $\gamma_-$ (stronger transitions between the $m_s$ = 0 and -1 states, solid) and $\gamma_+$ (weaker transitions between the $m_s$ = 0 and 1 states, dashed). **c** Optically detected magnetic resonance (ODMR) of the hyperfine electronic-spin transitions. See main text for discussion of the data. Inset: ODMR pulse sequence – 532nm laser pulses initialize and readout the electronic spin state, a MW $\pi$-pulse drives transitions, and an avalanche photodiode (APD) collects photons. **d** Spin-flip probability in a single optical cycle for $\gamma_\pm$. Near the excited state level avoided crossing (orange dashed line), $\gamma_-$ is much stronger than $\gamma_+$. Inset: The shift of electronic-spin levels with magnetic field $B_0$, showing the crystal field splitting parameter $D_{es}/h = 1.42$ GHz in the excited state and the level anti-crossing (orange circle).

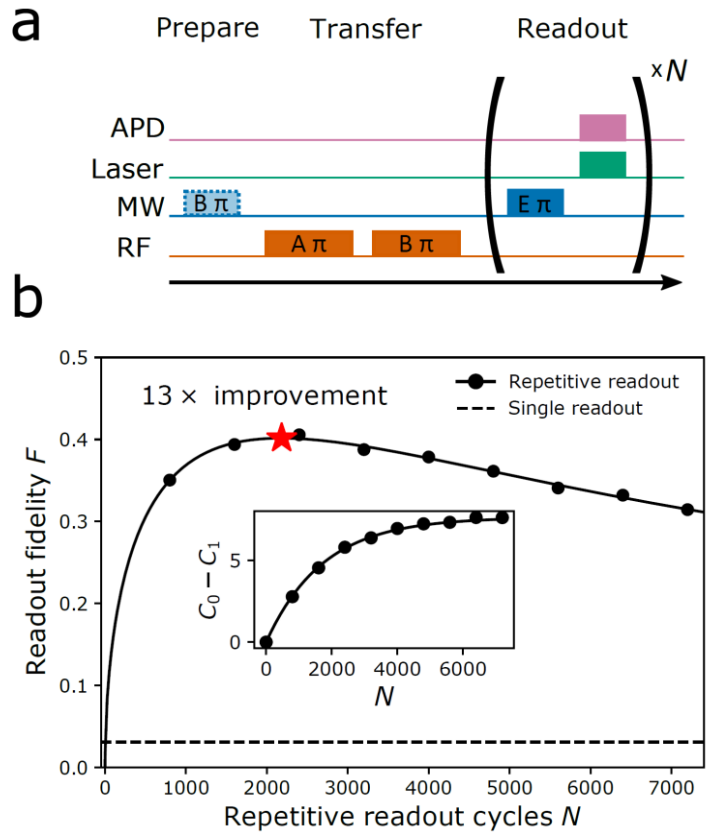

**Figure 2** Repetitive readout of a nuclear-spin memory in a nanodiamond. **a** Repetitive readout characterization protocol. After the nuclear spin is initialized (not shown), the electronic spin is prepared in either $|0\rangle_e$ or $|-1\rangle_e$ via the MWB transition and the electronic-spin state is stored in the nuclear spin by a CNOT gate composed of two RF pulses. Finally, the nuclear spin state is repetitively readout $N$ times. **b** Readout fidelity of the repetitive readout protocol at $B_0 = 244\ mT$. Error bars denoting standard error are smaller than the data markers. Inset: Difference in number of counts per measurement $C_0$ (preparation in $|0\rangle_e$) and $C_1$ (preparation in $|-1\rangle_e$) for different numbers of readout cycles $N$.

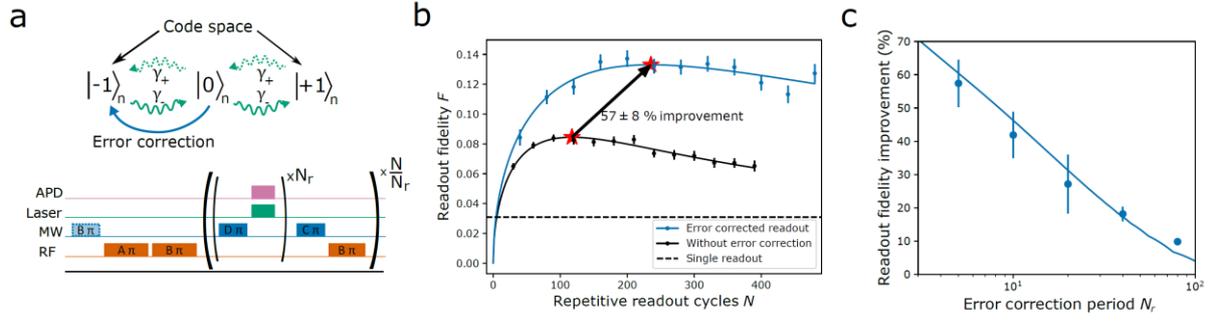

**Figure 3** Error-corrected repetitive readout. **a** Top: Illustration of the error correction protocol. The nuclear-spin code space states $\{|\pm 1\rangle_n\}$ store the electronic-spin state information. During readout, spin flip-flop processes induce memory errors on the $\gamma_+$ and $\gamma_-$ transitions (curvy green arrows). Errors that increment $m_I$ dominate the total error rate, and they can be corrected by periodic repumping (blue arrow), i.e. directional state transfer from the error state $|0\rangle_n$ back into the code state $|-1\rangle_n$. Bottom: Our implementation of the scheme, in which we perform the error correction pulses (a SWAP gate formed of MWC and RFB $\pi$-pulses) after an error-correction period of $N_r$ readout steps. **b** Comparison of the readout fidelity for protocols with and without error correction, when using $N_r = 5$ at a magnetic field $B_0 = 82\ mT$. Error bars show one-sigma standard errors. **c** The effect of the error-correction period $N_r$ on the readout fidelity of the error corrected readout scheme at a magnetic field $B_0 = 82\ mT$. The solid line shows the results of a master-equation model for the population dynamics (see text for details). Error bars show one-sigma fitting errors.

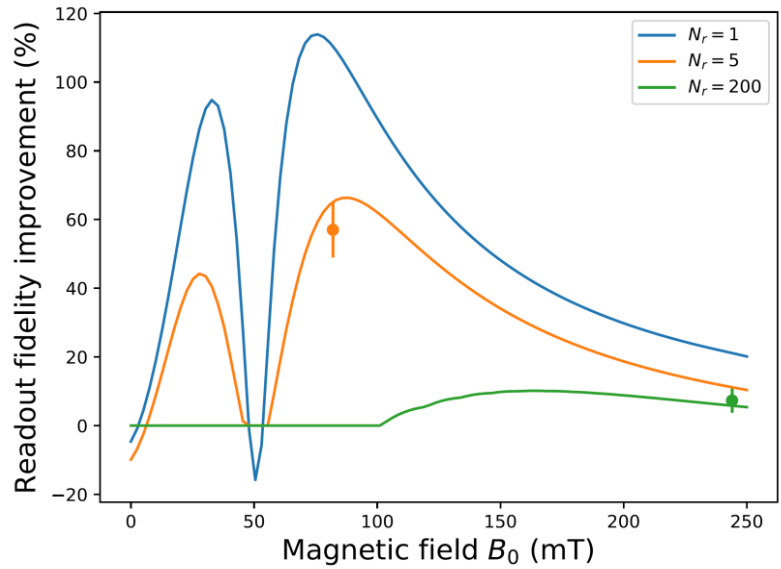

**Figure 4** Error-corrected readout performance under different magnetic field conditions. The improvement in repetitive readout fidelity from error correction is shown as a function of applied magnetic field $B_0$ for different error-correction periods of $N_r$ readout steps. The solid lines show the improvement calculated using the master equation model. Measured improvements are plotted as points, with error bars corresponding to the standard fit error.